\documentclass[prl,twocolumn,superscriptaddress,showpacs]{revtex4}
\usepackage{graphicx}
\usepackage{epsfig}
\usepackage{amssymb,amsmath,amsfonts,hyperref}
\usepackage{wasysym}
\usepackage{latexsym}
\usepackage{eucal}
\usepackage{verbatim}
\usepackage[normalem]{ulem}

\usepackage{color}

\newcommand{\be}{\begin{equation}}
\newcommand{\ee}{\end{equation}}

\newcommand{\bea}{\begin{eqnarray}}
\newcommand{\eea}{\end{eqnarray}}
\newcommand{\ra}{\rangle}
\newcommand{\la}{\langle}

\begin{document}

\title{N\'eel to valence-bond solid transition on the honeycomb lattice: Evidence for deconfined criticality}

\author{Sumiran Pujari}
\affiliation{Laboratoire de Physique Th\'eorique, Universit\'e de Toulouse and CNRS, UPS (IRSAMC), F-31062 Toulouse, France}
\author{Kedar Damle}
\affiliation{Department of Theoretical Physics, Tata Institute of Fundamental Research, Mumbai 400 005, India}
\author{Fabien Alet}
\affiliation{Laboratoire de Physique Th\'eorique, Universit\'e de Toulouse and CNRS, UPS (IRSAMC), F-31062 Toulouse, France}

\begin{abstract}
We study a spin-$1/2$ $SU(2)$ model on the honeycomb lattice with nearest-neighbor antiferromagnetic
exchange $J$ that favors N\'eel order, and competing 6-spin interactions $Q$ which favor a valence bond solid (VBS) state in which the bond-energies order at the ``columnar'' wavevector ${\mathbf K} = (2\pi/3,-2\pi/3)$. We present
quantum Monte-Carlo evidence for a direct continuous quantum phase transition between N\'eel and VBS states, with exponents and logarithmic violations
of scaling consistent with those at analogous deconfined critical points on the square lattice. Although this strongly suggests a description in terms
of deconfined criticality, the measured three-fold anisotropy of the phase of the VBS order parameter shows unusual near-marginal behaviour at the critical point.

\end{abstract}
\maketitle

Many interesting
materials at low temperature appear to be on the verge of
a quantum phase transition involving a qualitative change in
the nature of the ground state\cite{quantumcriticalityreview}. 
When one of the two competing $T=0$ phases spontaneously breaks
a symmetry, the transition can be studied
using a path integral representation with a Landau-Ginzburg action\cite{Landau} written in terms
of the order parameter that characterizes the broken
symmetry phase\cite{quantumcriticalityreview}.
If phases on two sides of the critical point break different symmetries, Landau-Ginzburg theory generically predicts a direct first-order transition
or a two-step transition with an intermediate phase.
However, this path integral description in terms of order-parameter variables
can sometimes involve Berry phases in a non-trivial way\cite{Haldane,Read_Sachdev_PRL89,Read_Sachdev_PRB90}. The presence of
Berry phases, which correspond to complex Boltzmann weights for the corresponding classical statistical mechanics problem in one higher dimension\cite{quantumcriticalityreview}, can invalidate the conclusions reached
by the Landau-Ginzburg approach.

In some of these cases, it is useful\cite{Levin_Senthil} to think in terms of topological
defects in one of the ordered states, and view the competing ordered state
as being the result of the condensation of these topological defects --- this
description\cite{Levin_Senthil} makes sense only if the quantum numbers carried
by defects in one phase match those of the order parameter variable
in the other phase. Under certain conditions, this alternate ``non-Landau'' description
generically predicts a direct continuous transition\cite{Senthil_etal_Science,Senthil_etal_PRB} between the two ordered
states, in contrast to predictions of classical Landau-Ginzburg theory. 
Square lattice $S=1/2$ antiferromagnets undergoing a transition from
a ground state with non-zero N\'eel order parameter $\vec{M}_s$
to a valence-bond solid (VBS)
ordered state, in which the ``bond-energies'' (singlet
projectors) $P_{\langle ij\rangle}
\equiv \frac{1}{4} - \vec{S}_i \cdot \vec{S}_j$ on nearest-neighbour
bonds $\langle i j \rangle$  in the $\hat{x}$ ($\hat{y}$)
direction develop
long-range order at the ``columnar'' wavevectors ${\mathbf K}_1 = (\pi,0)$ (${\mathbf K}_2 = (0,\pi)$), provide the best-studied example of such
``deconfined critical points''\cite{Senthil_etal_PRB,Senthil_etal_Science}.
In this case, $Z_4$ vortices in the complex VBS order parameter $\Psi$ carry a net spin $S=1/2$ in their core, suggesting that the onset of N\'eel order
can be studied using a CP$^1$ description of $\vec{M}_s$: $\vec{M}_s =  z^{*}_{\alpha} \vec{\sigma}_{\alpha
\beta} z_{\beta}$, where $\vec{\sigma}$ are Pauli matrices
and the $Z_4$ vortices are represented by a two-component complex bosonic field $z_{\alpha}$ coupled
to a compact U(1) gauge field ${\mathcal A}_\mu$\cite{Levin_Senthil,Senthil_etal_Science,Senthil_etal_PRB}, whose
space-time monopoles correspond\cite{Read_Sachdev_PRL89,Dadda} to hedgehog defects in the N\'eel order.
Only quadrupled hedgehog defects (corresponding to four-fold anisotropy in the phase of $\Psi$)  survive
the destructive interference of Berry phases on the square lattice\cite{Haldane,Read_Sachdev_PRL89,Read_Sachdev_PRB90,Dadda}, and their irrelevance
at criticality\cite{Senthil_etal_PRB,Senthil_etal_Science} leads to a {\em non-compact} (monopole-free \cite{Lau_Dasgupta, Kamal_Murthy, Motrunich_Vishwanath}) CP$^1$ (NCCP$^1$) description of this transition.

\begin{figure}
{\includegraphics[width=\hsize]{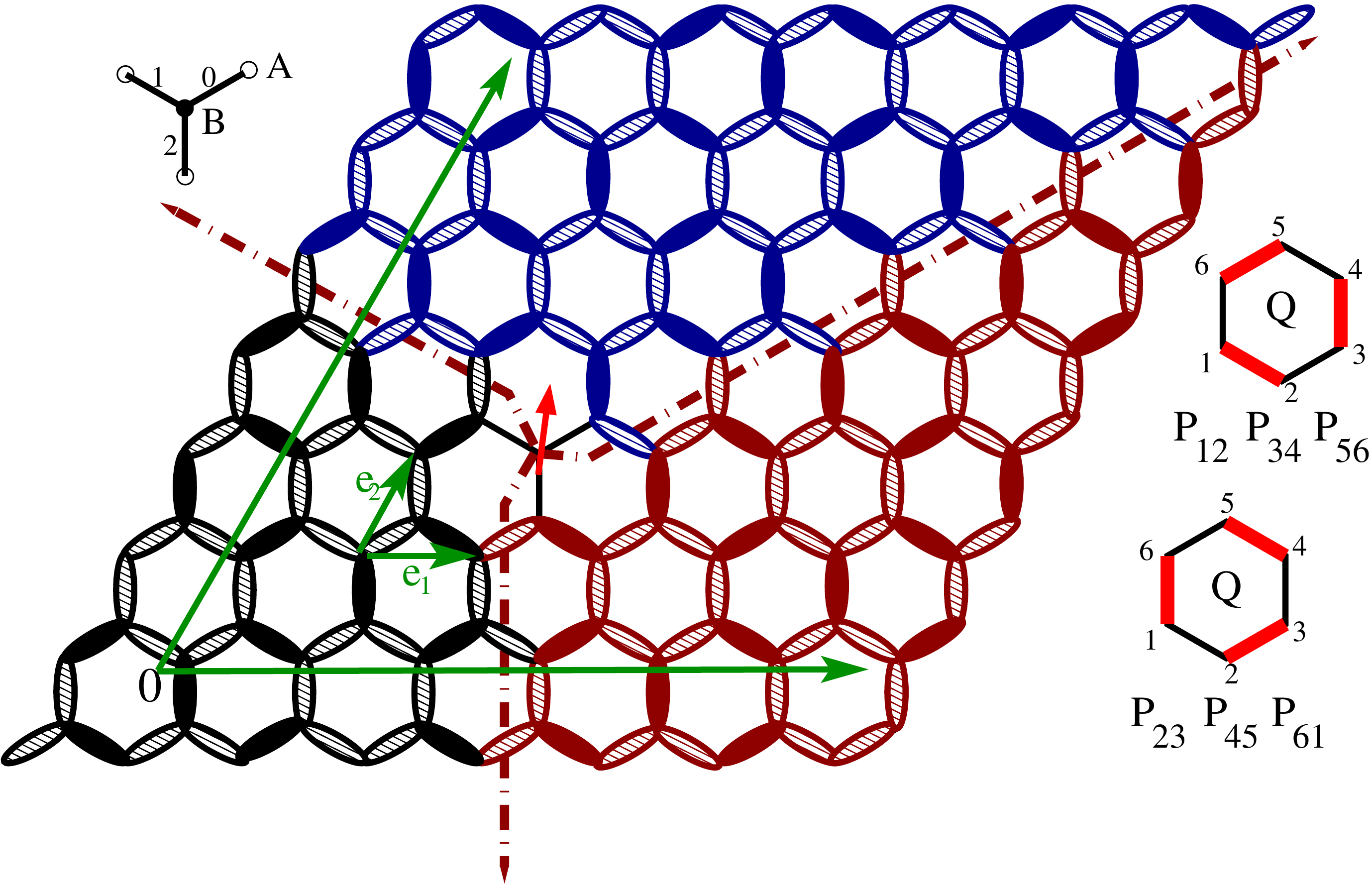}}
\caption{(Color online) The honeycomb lattice has a two-site
basis (labeled $A$ and $B$)
and elementary Bravais lattice
translations $\hat{e}_1$ and $\hat{e}_2$, with distances from origin specified in units of $\hat{e}_1$ and $\hat{e}_2$. Three types of bonds (labeled
$0$, $1$, $2$), oriented
along the three principal directions, ``belong'' to each Bravais lattice site.
Columnar and plaquette VBS order at wavevector ${\bf K}$ correspond to different choices of the order
parameter phase, with solid filled dimers on a link $\langle i j \rangle$
denoting high (low) values of $\langle P_{\langle i j \rangle} \rangle$ in the columnar (plaquette) state.
Here three domain
walls meet at the core of the $Z_3$ vortex, which carries
a net $S=1/2$ spin. Also shown is a depiction of the six-spin
interaction terms in Eq. 1.
}
\label{Fig1}
\end{figure}

Here, we use Quantum Monte Carlo (QMC) simulations\cite{Sandvik_PRL2005,Sandvik_Evertz_PRB2010,SSE} to study a spin-$1/2$ Heisenberg model on the honeycomb lattice with nearest-neighbor antiferromagnetic
exchange $J$ that favors N\'eel order, and competing 6-spin interactions $Q$ which favor VBS order at the columnar wavevector ${\mathbf K} = (2\pi/3,-2\pi/3)$:
\begin{equation*}
H = -J \sum_{\la ij \ra} P_{\langle ij \rangle} -Q \sum_{\la \la ijklmn\ra \ra}
\!\!\!\!\!\!\!\!(P_{\langle ij \rangle} P_{\langle kl \rangle} P_{\langle mn \rangle} + P_{\langle jk \rangle}P_{\langle lm \rangle}P_{\langle ni \rangle}) \;,
\label{eq:JQ}
\end{equation*}
where $\la \la ijklmn \ra \ra$ denotes hexagonal plaquettes (Fig~\ref{Fig1}). We find evidence for a direct continuous N\'eel-VBS transition
at  $(Q/J)_c \equiv q_c \approx 1.190(6)$, with correlation length exponent $\nu \approx 0.54(5)$, and anomalous exponents $\eta_{\rm Neel} \approx 0.30(5)$, and $\eta_{\rm VBS} \approx 0.28(8)$; within errors, these values match corresponding results at the N\'eel-columnar VBS transition on the square lattice\cite{Sandvik_PRB2012,Sandvik_PRL2010,Sandvik_PRL2007}. In addition, we find evidence for apparently logarithmic violations of finite-temperature scaling of the uniform spin susceptibility $\chi_u$ and stiffness $\rho_s$, analogous to the square-lattice case\cite{Sandvik_PRL2010}. However, in sharp contrast
to the square-lattice transition at which the {\em four-fold} anisotropy vanishes for 
large systems\cite{Sandvik_PRL2007,Lou_Sandvik,Lou_etal_PRB09}, a careful study of the {\em three-fold} anisotropy in the phase of $\Psi$
reveals surprising near-marginal behaviour on the honeycomb
lattice.

To put these results in context, we first note that $Z_3$ vortices in $\Psi$ carry a net spin $S=1/2$ in their core on the honeycomb lattice (Fig.~\ref{Fig1}) analogous to $Z_4$ vortices on the square lattice. Therefore, a continuum CP$^{1}$ description\cite{Levin_Senthil} is
again appropriate. 
The monopole creation operator in the CP$^{1}$ description
transforms under lattice symmetries
in {\em the same way} as the complex VBS order parameter $\Psi$ 
at the columnar wavevectors
on both the honeycomb and square lattices, allowing one to view these VBS states as monopole condensates\cite{Read_Sachdev_PRL89,Read_Sachdev_PRB90,Senthil_etal_PRB,Senthil_etal_Science}.
On the honeycomb lattice, it picks up a $2\pi/3$ phase under
lattice rotations. Therefore,
insertions of {\em tripled} monopoles are allowed
on the honeycomb lattice, and manifest themselves as a {\em three-fold}
anisotropy felt by the phase of $\Psi$. If this is relevant, one expects the
correct long-wavelength description of the transition to be a conventional Landau-Ginzburg theory written in terms of $\vec{M}_s$ and $\Psi$, and the transition to be first-order in the simplest scenario,
or proceed in two steps with an intermediate phase \cite{Senthil_etal_PRB,Senthil_etal_Science}. On the square lattice, 
only {\em quadrupled} monopoles are allowed in the CP$^1$
description since $\Psi$ picks up a $\pi/2$ phase under rotations. These can be straightforwardly argued\cite{Senthil_etal_PRB,Senthil_etal_Science} to be irrelevant in the NCCP$^{N-1}$ theory at $N=2$
by noting that they are irrelevant
{\em both} at $N=1$\cite{Read_Sachdev_PRB90,Senthil_etal_PRB,Sachdev_Jalabert,Oshikawa,Lou}, {\em and}
in the $N \rightarrow \infty$ limit\cite{Read_Sachdev_PRB90,Senthil_etal_PRB,Sachdev_Jalabert}, leading to a NCCP$^{1}$ 
description of the transition.
\begin{figure}
{\includegraphics[width=\hsize]{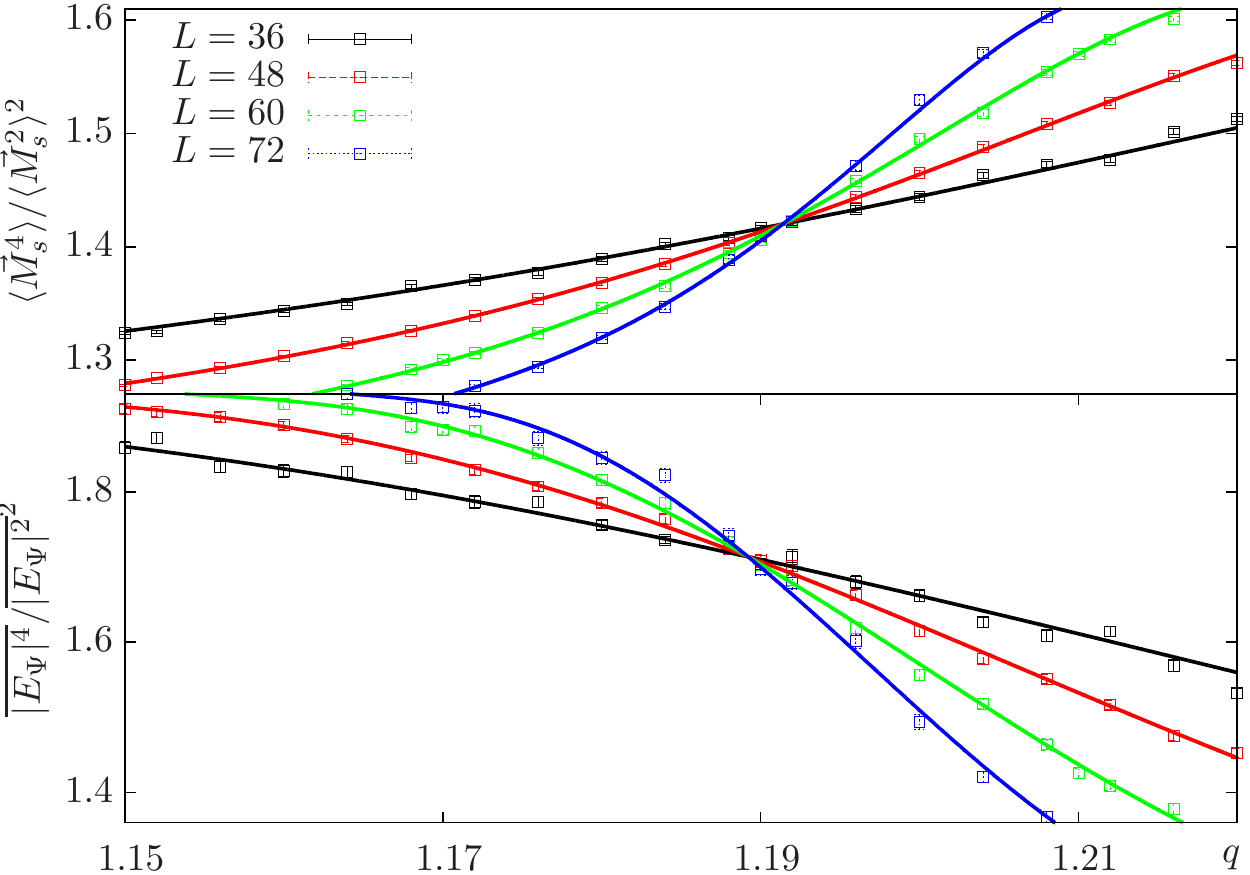}}
\caption{(Color online) Binder cumulants of $\vec{M}_s$ and $E_{\Psi}$ as a function of $q$ for different sizes $L$ (symbols), fit to a polynomial in $(q-q_{cD/N}).L^{1/\nu_{D/N}}$ (lines) with best-fit values $\nu_N=0.5080$, $\nu_{D}=0.5237$, $q_{cN}=1.1912$ and $q_{CD}=1.1892$. Best-fit values are for the $L\geq 48$ part of the displayed data.}
\label{Fig2}
\end{figure}

Thus, on one hand, the continuous nature and measured exponents of the honeycomb lattice transition, as well as the finite-temperature behaviour of $\rho_s$ and $\chi_u$, point
to a NCCP$^1$ description and suggest that {\it tripled} monopoles are
irrelevant at the NCCP$^1$ fixed point,
allowing the physics of deconfined criticality to control universal properties of transitions to VBS order at
wavevector ${\bf K}$.
If the transition to plaquette VBS order at the {\em same} wavevector ${\bf K}$ (Fig. 1) 
in the frustrated $J_1$-$J_2$
is indeed direct and continuous\cite{Albuquerque,Ganesh,Zhu,Gong}, our results suggest, 
on grounds of universality, that it too would be governed by the NCCP$^1$ fixed point.
On the other hand, our observation of near-marginal behaviour of
the three-fold anisotropy at criticality suggests that three-fold monopoles
remain important ingredients of the honeycomb lattice transition at large
scales, making it remarkable that other signatures of the transition
conform to what one expects at the NCCP$^1$ critical point.
The physical $N=2$ case lies between two contrasting extremes of the NCCP$^{N-1}$ theory: tripled monopoles are relevant at $N=1$\cite{Read_Sachdev_PRB90,Senthil_etal_PRB,Sachdev_Jalabert,Oshikawa} and lead to
a {\em weakly-first order transition}\cite{Janke}, but strongly irrelevant in the $N \rightarrow \infty$ limit\cite{Read_Sachdev_PRB90,Senthil_etal_PRB,Sachdev_Jalabert}. Our results
therefore suggest that tripled-monopoles switch from relevant
to irrelevant behaviour at or very close to $N=2$ as one increases $N$ in
the NCCP$^{N-1}$ theory. 
\begin{figure}
{\includegraphics[width=\hsize]{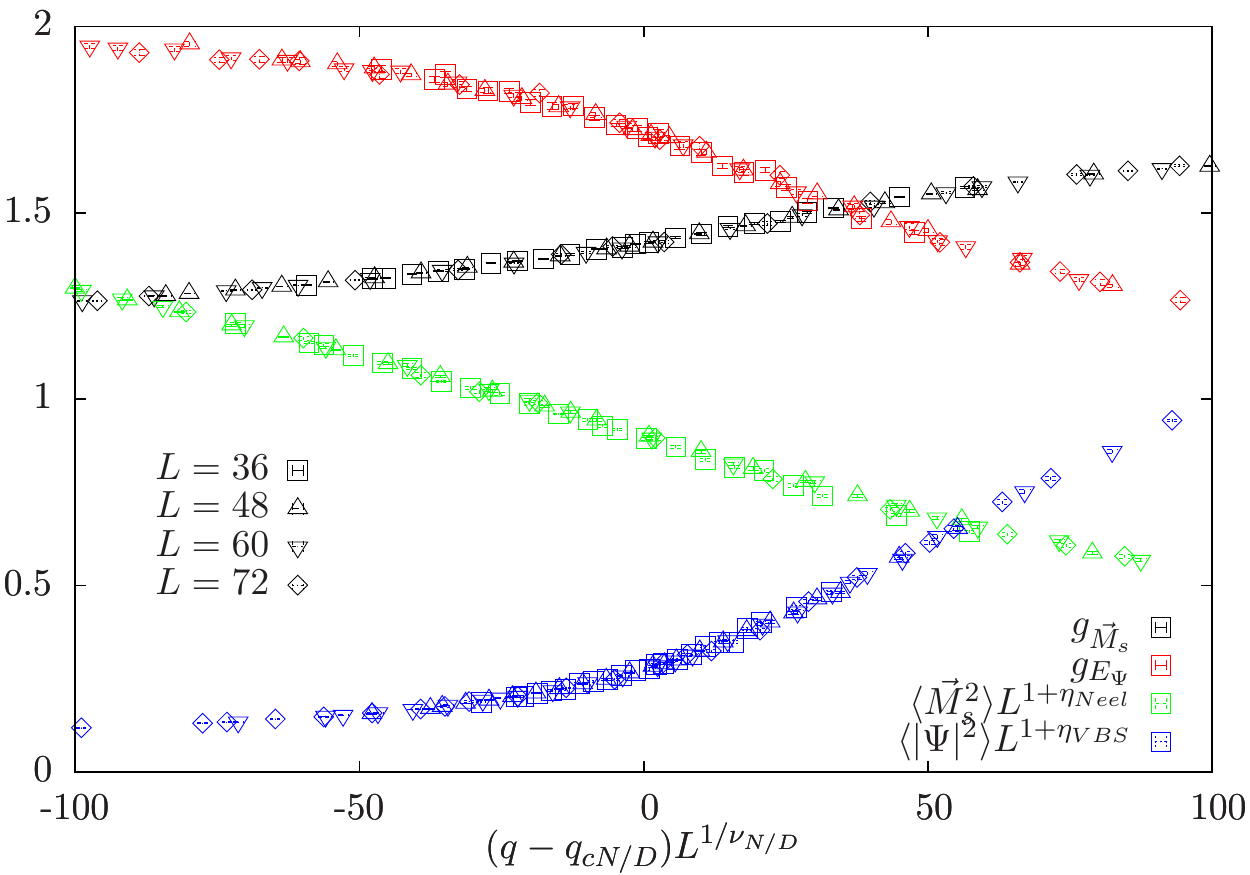}}
\caption{(Color online) Scaling collapse of Binder cumulants of $\vec{M}_s$ and 
$E_{\Psi}$, using values of $q_{cD}$, $q_{cN}$, $\nu_D$
and $\nu_N$ quoted in legend of Fig.~\ref{Fig2}. Similar collapses for $\langle \vec{M}_s^2\rangle L^{1+\eta_{\rm Neel}}$, $\langle |\Psi|^2\rangle L^{1+\eta_{\rm VBS}}$ are also displayed, obtained using the following best-fit values: $q_{cN} = 1.1956$, $\nu_N = 0.5003$, $\eta_{\rm Neel}=0.3539$ ($\langle \vec{M}_s^2\rangle$) and $q_{cD} = 1.1864$, $\nu_D = 0.558$, $\eta_{\rm VBS} = 0.25$ ($\langle |\Psi|^2\rangle$). Best-fit values are for the $L \geq 48$ part of the displayed data.}
\label{Fig3}
\end{figure}

\begin{figure}
\includegraphics[width=\hsize,angle=0,trim= 16 8 12 11, clip=true]{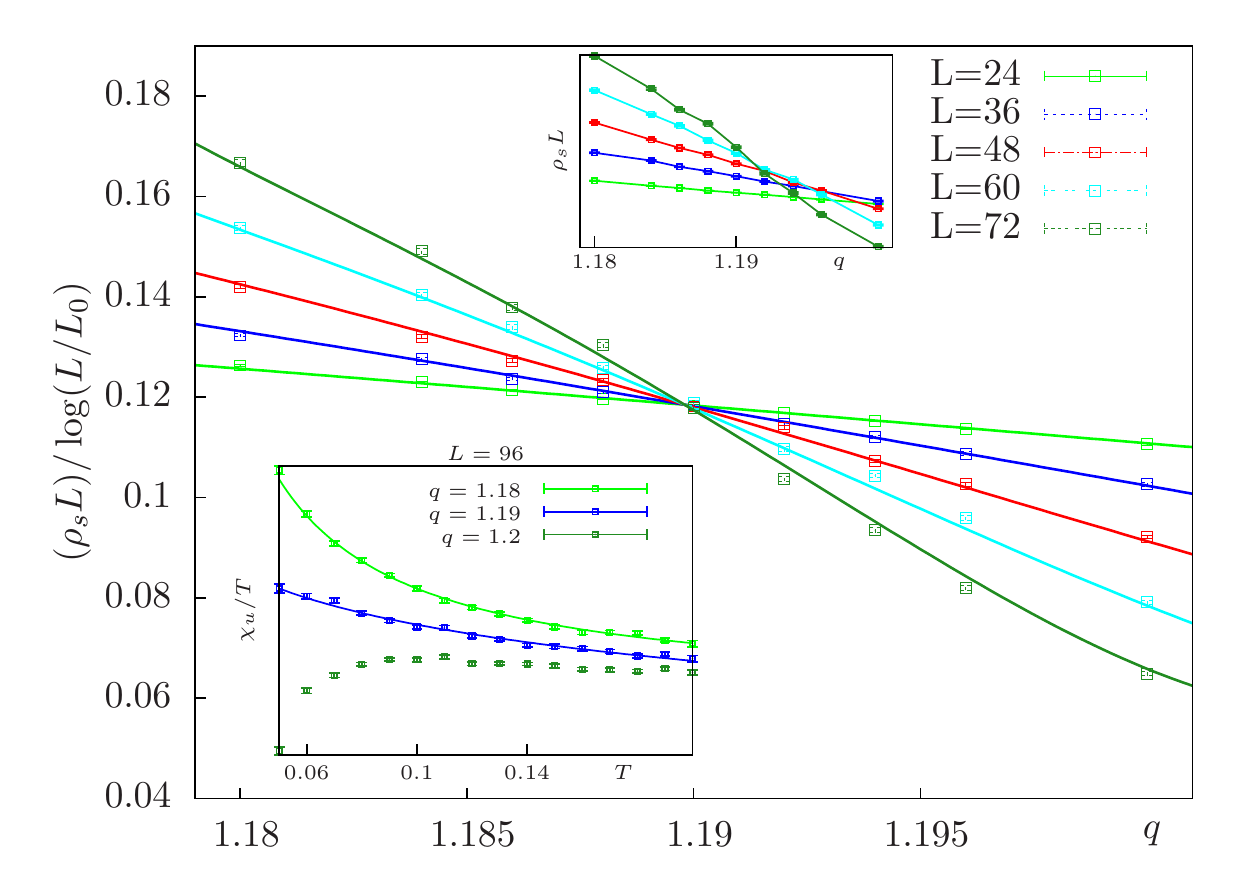}
\caption{(Color online) $\rho_sL$ does not obey
standard quantum-critical scaling $\rho_s L = h((q-q_{cN})L^{1/\nu_N})$ with dynamical exponent $z=1$
(for instance, see drift in $\beta=2L$ data shown in top inset). In contrast
$\rho_s L/\log(L/L_0)$ with $L_0 =0.37$ shows excellent scaling. Symbols are QMC data, and lines best-fit to this modified scaling form, with $q_{cN} \approx 1.190(2)$ and $\nu_N \approx 0.54(2)$ in agreement with our $T=0$ results. Bottom inset : Temperature dependence of $\chi_u / T$ close to criticality. In the N\'eel phase ($q=1.18$), QMC data (symbols) are well-fit by $\chi_u/T= a + b/T$, whereas on the VBS side ($q=1.2$), a sharp drop is observed as expected. Close to criticality ($q=1.19$), QMC data are better fit by $\chi_u/T = c + d\log(J/T)$. Lines are fits to the above forms with $a=0.024$, $b=0.0005$, $c=0.022$ and $d=0.0024$.}
\label{fig:log_corrections}
\end{figure}

It is quite clear that the continuous transition studied here is
very different from
transitions to {\em staggered VBS order}
on square and honeycomb lattices\cite{Banerjee_Damle_Paramekanti_2010,Sen_Sandvik}, whose strongly first order nature can be attributed\cite{Banerjee_Damle_Paramekanti_2010} to the spinless cores of vortices in staggered VBS states\cite{Levin_Senthil}. Indeed, most of our results on universal critical properties are very similar to previous QMC simulations of computationally tractable spin models exhibiting N\'eel-columnar VBS transitions on the square lattice\cite{Sandvik_PRB2012,Sandvik_PRL2010,Sandvik_PRL2007,Banerjee_etal_2010,Kaul_2011,Banerjee_etal_2011,Lou_etal_PRB09,Kaul_Sandvik_PRL2012,Melko_Kaul_PRL2008,Jiang_etal_JStatmech2008,Chen}. While some of these studies\cite{Sandvik_PRB2012,Sandvik_PRL2007,Sandvik_PRL2010,Banerjee_etal_2010,Kaul_2011,Banerjee_etal_2011,Lou_etal_PRB09,Kaul_Sandvik_PRL2012,Melko_Kaul_PRL2008} have interpreted these square-lattice results
within the framework of the NCCP$^1$ theory, albeit with some
logarithmic violations of scaling\cite{Sandvik_PRL2010,Banerjee_etal_2010,Kaul_2011,Banerjee_etal_2011}, other studies\cite{Jiang_etal_JStatmech2008,Chen} have
interpreted very similar numerical data in terms of a flow to a very weakly
first order transition at large length-scales---this is motivated
by data on lattice-discretized NCCP$^1$ models\cite{Motrunich_Vishwanath2,Kuklov}, some of which exhibit first-order behaviour\cite{Kuklov}. Our work adds another dimension
to this debate by demonstrating that results otherwise consistent with
the NCCP$^1$ description are accompanied by significant anisotropy
in the phase of $\Psi$ at the honeycomb lattice transition.
\begin{figure}
\includegraphics[width=\hsize,angle=0,trim= 0 0 0 0, clip=true]{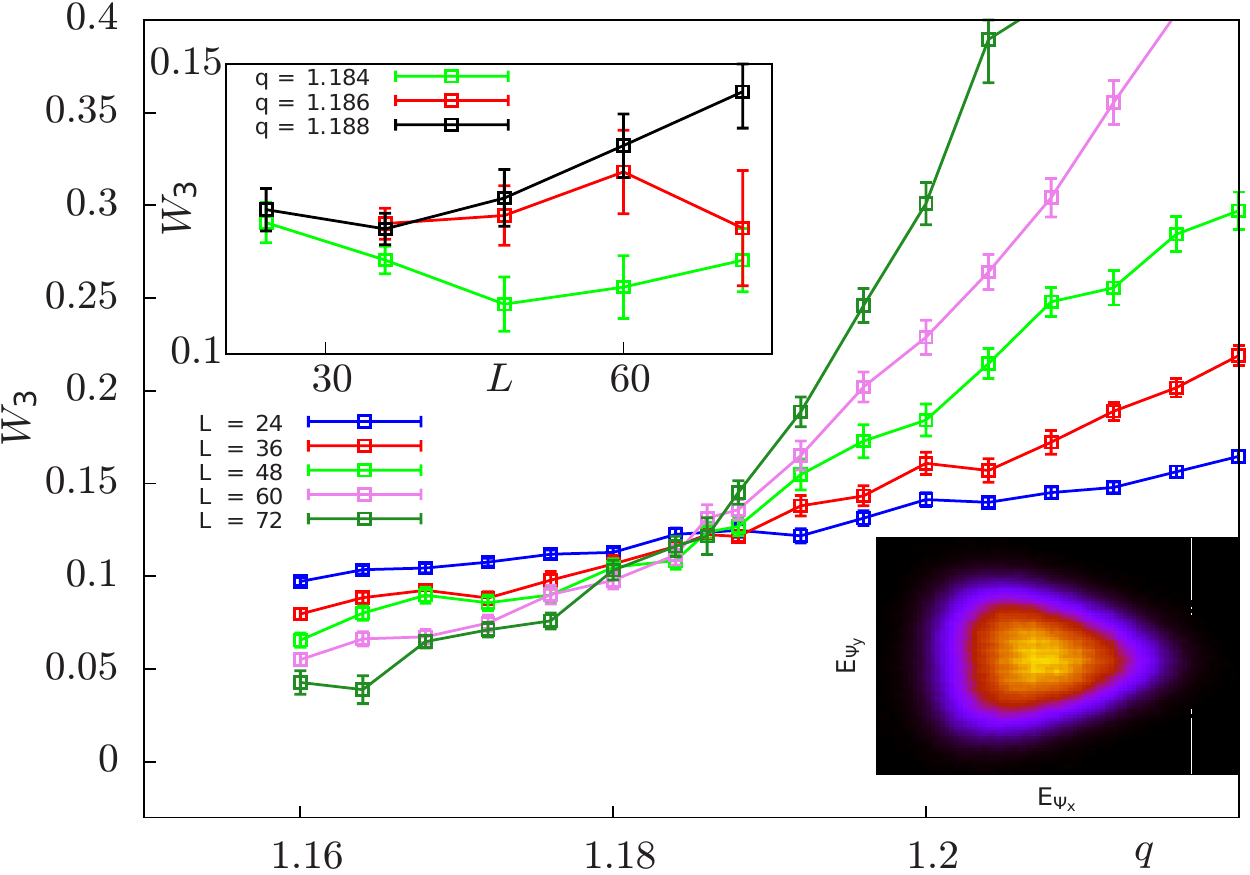}
\caption{(Color online) The dimensionless $Z_3$-anisotropy parameter $W_3$ scales to
zero with increasing $L$ in the N\'eel phase, but grows with size
in the columnar VBS phase. Top inset zooms in on behaviour of near-critical
systems which display nearly scale-independent behaviour. Bottom inset: Histogram of $E_{\Psi}$ for $L=36$ at $q=1.184$ close to
$q_{cD}$. Brightness of each color patch reflects the weight.} 
\label{fig:histogram}
\end{figure}

We study $H$ on $L \times L$
honeycomb lattices (Fig~\ref{Fig1}) of $2L^2$ spins, with periodic boundary conditions and $L$ a multiple of $12$ up to $L=72$. We use a $T=0$ projector QMC algorithm\cite{Sandvik_Evertz_PRB2010}, with
a sufficiently large projection length $cL^3$  ($c$ ranging from $4$ to $12$) to ensure convergence to the ground state.
At small $q$, the ground-state is N\'eel ordered, as characterized by the N\'eel order parameter $\vec{M}_s = \frac{1}{2L^2}\sum_{\vec{r}} \vec{m}(\vec{r})$,
where $\vec{m}$ is the local N\'eel field $\vec{m}(\vec{r}) = \vec{S}_{\vec{r} A} - \vec{S}_{\vec{r}B}$.
Here $\vec{r}A$ ($\vec{r}B$) refers to the $A$ ($B$) sublattice
site belonging to Bravais lattice site $\vec{r}$ (Fig.~\ref{Fig1}). To locate the quantum phase transition where N\'eel order is lost, we compute the ``dimensionless" Binder cumulant $g_{\vec{M}_s} = {\langle (\vec{M}_s^2)^2\rangle} / {\langle \vec{M}_s^2\rangle^2}$. It is expected to
obey a scaling form $F_{g_{\vec{M}_s}}(\Delta q_N)$ if there is a continuous
transition at $q_{cN}$. Here, $F_{g_{\vec{M}_s}}$ is a universal scaling function of the argument $\Delta q_N \equiv (q-q_{cN})L^{1/\nu_{N}}$
where $\nu_N$ is the correlation length exponent associated with N\'eel correlations. In the vicinity of such a transition, we also expect the scaling form $\langle \vec{M}_s^2\rangle = L^{-(1+\eta_{\rm Neel})}G_{\vec{M}_s}(\Delta q_N)$ for the corresponding dimensionful quantity.

At large $q$, we find that VBS order develops at the columnar wavevector ${\bf K}$. This is characterized by the VBS order parameter $\Psi = \frac{1}{2L^2}\sum_{\vec{r}} V_{\vec{r}}$, where $V_{\vec{r}}$ is the local VBS order parameter field:
\begin{equation*}
V_{\vec{r}} = (P_{\vec{r}0} + e^{2 \pi i /3} P_{\vec{r} 1} + e^{4 \pi i/3} P_{\vec{r} 2})e^{i {\mathbf K} \cdot \vec{r}}\; .
\end{equation*}
Here, $P_{\vec{r} \mu}$ ($\mu = 0, 1,2$) denotes the singlet projector on the bond $\mu$ ``belonging" to Bravais lattice site $\vec{r}$ (Fig.~\ref{Fig1}). To  quantify the strength of VBS order, we compute $\langle |\Psi|^2\rangle = \langle \Psi^{\dagger}\Psi\rangle$. The phase of $\Psi$ distinguishes
between two kinds (columnar vs plaquette) of three-fold symmetry breaking VBS order at wavevector ${\mathbf K}$. In the $T=0$ QMC simulations, information on
this phase is obtained from the estimator $E_{\Psi}$, whose average $\overline{E_{\Psi}}$ over the QMC run gives the quantum-mechanical expectation value $\langle \Psi\rangle$. Although $E_\Psi$  is a basis dependent quantity, the histogram of its phase can nevertheless be used to distinguish between the different VBS states at the same wavevector\cite{Sandvik_PRL2007,Lou_etal_PRB09}. 
The VBS transition can be located by focusing again on a dimensionless quantity, the (basis-dependent) Binder cumulant\cite{footer} of $E_\Psi$ defined as $g_{E_\Psi} \equiv \overline{|E_\Psi|^4}/\left(\overline{|E_{\Psi}|^2}\right)^2$, which is again expected to obey a scaling form $F_{g_{E_\Psi},D}(\Delta q_D)$ if VBS order is lost via a continuous $T=0$ transition at $q_{cD}$. The argument
 $\Delta q_D \equiv (q-q_{cD})L^{1/\nu_{D}}$ of the universal scaling function $F_{g_{E_\Psi},D}$ uses $\nu_D$, the correlation length exponent associated with
VBS correlations. Close to such a continuous transition, we also expect the corresponding scaling form
$\langle |\Psi|^2\rangle = L^{-(1+\eta_{\rm VBS})}G_{\Psi}(\Delta q_D)$ for the dimensionful observable.

We pinpoint the $T=0$ N\'eel and VBS transitions from the crossings
of the Binder ratios $g_{\vec{M}_s}$ and $g_{E_\Psi}$ as a function of $q$ for various $L$---at this stage, we do not assume that the two transitions coincide. Given the relatively sharp nature of the crossings and the monotonic nature of their $q$ dependence for fixed $L$ (Fig~\ref{Fig2}), we are confident that the transition(s) is (are) continuous. We fit data for each dimensionless  ($g_{\vec{M}_s}$, $g_{E_\Psi}$) and  (appropriately scaled) dimensionful quantity  $\langle \vec{M}_s^2\rangle.L^{1+\eta_{\rm Neel}}$, $\langle |\Psi|^2\rangle.L^{1+\eta_{\rm VBS}}$, in the critical range to a polynomial function of $(q-q_c)L^{1/\nu}$ (corresponding to a polynomial approximation of scaling functions), with the corresponding $q_c$, $\nu$, $\eta$ and polynomial coefficients being fitting parameters. For each dimensionless quantity, the best-fit values vary somewhat depending on the range of $L$ and $q$ studied. Results of such fits for one choice of data-set for the dimensionless quantities are displayed as lines in Fig.~\ref{Fig2}, with the corresponding scaling collapse displayed in Fig.~\ref{Fig3}. Similar results for N\'eel and VBS correlators \cite{supplementary} confirm
this.

Based on a detailed study of such fits, we estimate $q_{cN} \approx 1.1936(24)$, $q_{cD} \approx 1.1864(28)$,  $\nu_N=0.51(3)$, $\nu_D=0.55(4)$, $\eta_{\rm Neel} = 0.30(5)$ and $\eta_{\rm VBS} = 0.28(8)$. The error bars quoted here reflect not just the error in determining best-fit values for a given data-set
for each quantity, and variation in these best-fit values from quantity to quantity, but
also the dependence of these best-fit values on the data set used,
{\em i.e.} the size of the critical
window in $q$, and the range of $L$ used in the fits. We also emphasize that
our estimates of $\eta_{\rm VBS}$ and $\eta_{\rm Neel}$
depend sensitively on the value of $q_c$, resulting in the relatively large error bars quoted here. Nevertheless, we are in a position to exclude the relatively
tiny values of $\eta$ that characterize conventional second-order critical
points in $2+1$ dimensions.
Since $\nu_N$ {\em coincides} with $\nu_D$ within error bars,
and the allowed ranges of $q_{cN}$ and $q_{cD}$ almost touch at the one-sigma
level, the simplest interpretation of our data is that N\'eel order is lost and VBS order sets in at a single
continuous $T=0$ transition whose location is estimated to be $q_c \approx 1.190(6)$, with 
correlation exponent $\nu = 0.54(5)$, and anomalous exponents $\eta_{\rm Neel} = 0.30(5)$ and $\eta_{\rm VBS} = 0.28(8)$. This, taken together with the relatively large values of $\eta_{\rm Neel}$
and $\eta_{\rm VBS}$ characteristic of deconfined critical points, suggests an interpretation 
in terms of deconfined criticality. 

Indeed, our estimates of $\eta_{\rm VBS}$, $\eta_{\rm Neel}$, and $\nu$, as well as of the universal critical value $g^{*}=1.42(1)$ of the N\'eel Binder ratio at the $T=0$ transition are consistent within errors
with values for the analogous transition on the square lattice\cite{Sandvik_PRB2012,Sandvik_PRL2007,Sandvik_PRL2010}.
We also study the temperature dependence of the uniform spin susceptibility
$\chi_u$ and the antiferromagnetic spin stiffness $\rho_s$ using
finite-$T$ QMC methods\cite{SSE} at low temperatures
in the vicinity of this $T=0$ transition. As is clear from Fig. \ref{fig:log_corrections}, data for these quantities do not fit well to standard
scaling predictions. However, excellent data collapse is obtained
upon inclusion of logarithmic violations of scaling, using
the same functional forms employed earlier on the square lattice\cite{Sandvik_PRL2010}. These logarithmic violations
may be related to (near) marginal operators in the NCCP$^1$ theory
itself\cite{Nogueira_Sudbo,functionalRG}.

Finally, we turn to a study of the effective three-fold anisotropy felt
by the phase of $\Psi$ at criticality, as seen
in histograms of $E_{\Psi}$ near $q_c$. The phase $\theta$ of $E_{\Psi}$ (inset of Fig.~\ref{fig:histogram}) appears to
feel significant anisotropy near the $T=0$ transition on the honeycomb
lattice. To quantify this anisotropy in the distribution $P(E_{\Psi})$ near the critical point, we use a (dimensionless) estimator $W_3= \int d E_{\Psi} P(E_{\Psi}) \cos (3\theta)$, designed to be $0$ for a $U(1)$-symmetric distribution and $1$ ($-1$) for
ideal columnar (plaquette) VBS states (Fig.~\ref{Fig1}).
In Fig.~\ref{fig:histogram}, we see that
$W_3$ appears to saturate to a scale-independent constant at large $L$ as the transition is approached from the N\'eel phase, before growing with
size as one moves into a columnar VBS state. This near-marginal behaviour
of the anisotropy in $P(E_{\Psi})$ at the largest
scales accessible to our simulations is {\em very} different from the $U(1)$ symmetric probability distribution of $E_{\Psi}$
seen near the square lattice critical point\cite{Sandvik_PRL2007,Lou_Sandvik}.
A more refined scaling analysis \cite{supplementary} yields the same result,
leading us to our earlier suggestion that three-fold monopole insertions are (very close to) marginal at the NCCP$^1$ critical point---this is consistent
with recent parallel work that discusses relevance of $q$-fold
monopoles in SU($N$) spin models\cite{Block_Melko_Kaul,Harada_etal}.

We thank A.~Banerjee, L. Balents, S.~Capponi, R.~Kaul, A.~L\"auchli, M.~Mambrini, A.~Paramekanti and P. Pujol for useful discussions, and D.~Dhar and S.~N.~Majumdar for a critical reading of a previous version of our manuscript. 
We acknowledge computational resources from TIFR Mumbai, GENCI-CCRT (Grant 2012050225) and CALMIP. This research is supported by the Indo-French Centre for the Promotion of Advanced Research (IFCPAR/CEFIPRA) under Project 4504-1, French ANR program ANR-08-JCJC-0056-01, and in part by the National Science Foundation under Grant No. NSF PHY11-25915, during a visit by one of us (KD) to KITP, Santa Barbara. We also acknowledge support for collaborative visits from the University Paul Sabatier, Toulouse (KD) and TIFR (SP).


\begin{thebibliography}{999}
\bibitem{footer} We use $g_{E_\Psi}$ instead of the Binder cumulant $\langle |\Psi|^4\rangle/ \langle |\Psi|^2\rangle^2$ due to technical difficulties in measuring $\langle |\Psi|^4\rangle$.
\end{thebibliography}
\end{document}